# Shock Tube Design for High Intensity Blast Waves for Laboratory Testing of Armor and Combat Materiel

Elijah Courtney, Amy Courtney, Michael Courtney
BTG Research, 9574 Simon Lebleu Road, Lake Charles, LA, 70607
michael_courtney@alum.mit.edu

**ABSTRACT**
Shock tubes create simulated blast waves which can be directed and measured to study blast wave effects under laboratory conditions. It is desirable to increase available peak pressure from ~1 MPa to ~5 MPa to simulate closer blast sources and facilitate development and testing of personal and vehicle armors. Three methods were investigated to increase peak simulated blast pressure produced by an oxy-acetylene driven shock tube while maintaining suitability for laboratory studies. The first method is the addition of a Shchelkin spiral priming section which works by increasing the turbulent flow of the deflagration wave, thus increasing its speed and pressure. This approach increased the average peak pressure from 1.17 MPa to 5.33 MPa while maintaining a relevant pressure-time curve (Friedlander waveform). The second method is a bottleneck between the driving and driven sections. Coupling a 79 mm diameter driving section to a 53 mm driven section increased the peak pressure from 1.17 MPa to 2.25 MPa. Using a 103 mm driving section increased peak pressure to 2.64 MPa. The third method, adding solid fuel to the driving section with the oxy-acetylene, resulted in a peak pressure increase to 1.70 MPa.

1.0 INTRODUCTION

The use of improvised explosive devices (IEDs) has greatly increased in recent military conflicts, and as a direct result, more soldiers are also being exposed to explosions [1,2]. It has been shown that the blast wave from an explosion can cause injuries apart from projectiles or impacts; these have been called primary blast injuries. The recent increase in injury to personnel and blast-induced damage to materiel has motivated laboratory scale experiments on the effects of blast waves [3,4,5]. Goals of such experiments include improving armor and the treatment of blast-induced injuries. The shock tube is an instrument that is used to simulate a blast wave so that the simulated blast wave can be directed and measured more easily, and so experiments can be conducted in laboratory conditions [5,6].

Shock tubes have been used for over a century to study high speed aerodynamics and shock wave characteristics as well as the response of material to blast loading [7]. Relatively recently, the value of using shock tubes to understand and prevent blast-related injuries has been demonstrated. Most shock tube designs are one of two main categories based on how the simulated blast wave is created: compression-driven [8,9] or blast-driven [10,11]. However, each of these has limitations. Compression-driven shock tube designs often produce significant shot to shot variations in peak pressure, as well as pressure wave durations that are longer than those of realistic threats such as mines, hand grenades, and IEDs. Often, they do not approximate the Friedlander waveform of free field blast waves [12]. Furthermore, the expansion of the compressed gases results in a jet of expanding gases that transfers additional momentum to the test object. Blast-driven shock tubes produce more realistic profiles, but their operation requires expensive facilities, liability, and personnel overhead for storing and using high explosives [12].

Previous work showed that a modular, oxy-acetylene based shock tube produced realistic blast waves with peak pressures up to about 1.17 MPa [12]. However, in some situations it may be desirable to increase the peak pressure to as much as 5 MPa to simulate closer proximity to a blast source and assist development and testing of personal and vehicle armors. Higher blast pressures are also desirable for testing damage thresholds of equipment.

The present study investigated three approaches to increasing the peak pressure of the simulated blast wave produced with a laboratory scale oxy-acetylene based shock tube. The first method employs the addition of a Shchelkin spiral priming section which works by increasing the turbulent flow of the deflagration wave, thus increasing its speed and pressure. The second method uses a bottleneck between the driving and driven sections to increase pressure by increasing the ratio between volume of fuel and cross-sectional area of the driven section. The third method adds solid fuel to the driving section with the oxy-acetylene with the goal of increasing the heat and pressure of the blast wave inside the driven section.

2.0 MATERIALS AND METHODS

In all three designs, a single layer of food-grade plastic film (low density polyethylene) was placed over the open end to contain the mixture before filling the driving section with the fuel-oxygen mixture, with a small ventilation tube placed parallel to the driving section to allow ambient air to escape during filling. Two layers of Teflon tape were applied to the threads of the driving section before and after placement of the plastic film barrier to prevent the threads from cutting the film prior to ignition. Both driven and driving sections were commercially available steel pipe, and the sections were coupled by a steel flange. For the bottleneck and solid fuel designs, the driving section was sealed with a steel end cap, into which a hole was drilled for ignition access and the driving section was filled with a stoichiometric mixture of oxygen and acetylene. Combustion products of this mixture were carbon dioxide and water vapor. The ignition source, an electric match, was placed in the ignition access, which was then sealed with putty. The driving section was then threaded into the flange and the leads to the ignition source were attached to a remote nine volt DC source [12].

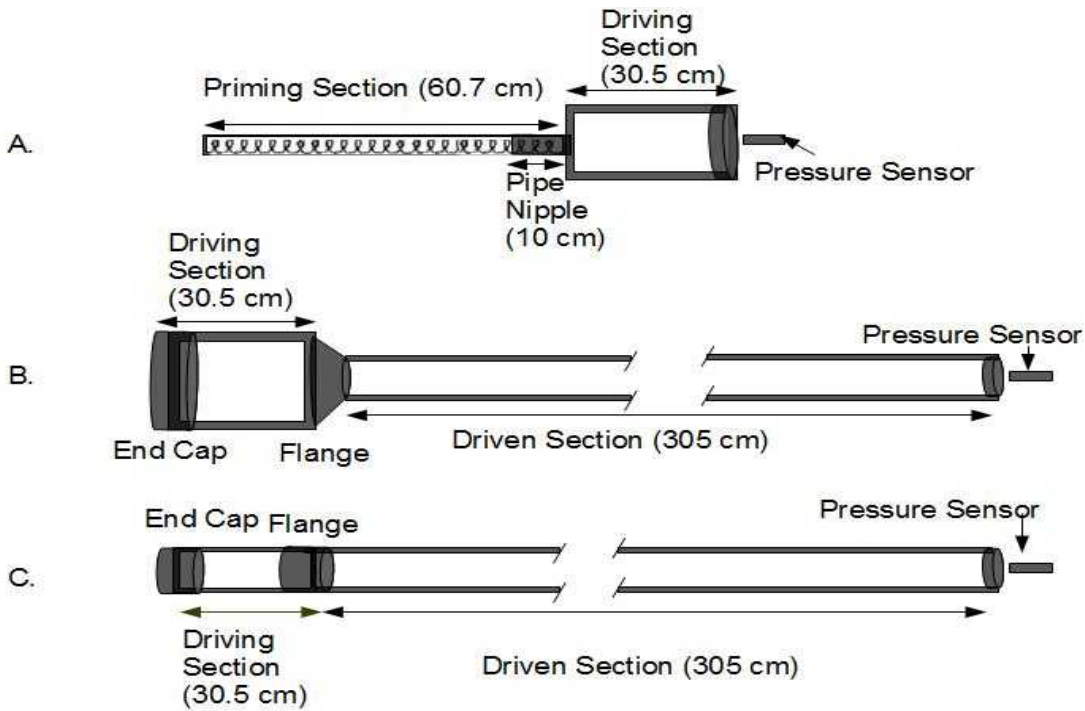

*Figure 1: Diagram of oxy-acetylene shock tube designs. A: Shchelkin spiral shock tube B: Bottleneck shock tube. C: Solid fuel shock tube.*

*2.1 Shchelkin spiral priming section* A Shchelkin spiral was incorporated into a priming section, which was placed behind the driving section (Figure 1A). The Shchelkin spiral is thought to work by increasing the turbulent flow of the deflagration wave, thus increasing the chemical reaction rate and wave speed [13,14]. Both the priming and the driving sections were filled with oxy-acetylene. For this design, the priming section was a 60.7 cm long 16 mm inner diameter machined steel tube with the spiral groove machined to a depth of 0.36 mm on the inside of the tube. The driving section was 30.5 cm long and 79 mm inner diameter. This design did not employ a driven section. The reaction of the priming compound (0.04 g of lead styphnate) was initiated by impact, thus igniting the oxy-acetylene. As the fuel burned along the priming section, the turbulence caused by the spiral supported a deflagration to detonation transition (a DDT). When the reaction reached the driving section, the energy was amplified by the additional volume of fuel in the driving section.

*2.2 Bottleneck driving section*
     A decrease in diameter from the driving section to the driven section, or a bottleneck, increased the ratio of chemical energy to the diameter of the driven section, while keeping other features of an earlier, successful design (Figure 1B). Two specific variations were tested in the present study: a 79 mm inner diameter 30.5 cm long steel cylinder as the driving section with a 53 mm inner diameter 304.8 cm long steel cylinder as the driven section; and a 103 mm diameter driving section with driven section of the same diameter. The expected pressure increase was by a factor of 2.1 for the 79 mm bottleneck, and a factor of 3.8 for the 103 mm diameter bottleneck, based on the assumption of proportionality to the increased volume of fuel [15].

*2.3 Addition of solid fuel*

An amount of additional solid fuel computed to yield 3 times the energy of the oxy-acetylene mix was added to the driving section in addition to the oxy-acetylene. The powdered solid fuel was ignited by the burning oxy-acetylene to try to increase the peak pressure by increasing the amount of chemical energy available. Calculations showed that 0.8 g of nitrocellulose was needed to triple the chemical energy. The driving section was 53 mm inner diameter and 30.5 cm long. The driven section was 53 mm inner diameter and 304.8 cm long (Figure 1C).

*2.4 Instrumentation*

For tests of each design, a piezoelectric pressure sensor (PCB Piezotronics 113B24) was placed at the shock tube opening with its face perpendicular to the direction of travel of the blast wave. Pressure data was recorded at a sample rate of 1 MHz via cables which connected the pressure transducer to a signal conditioning unit (PCB 842C) which produced a voltage output, which was digitized with a National Instruments USB-5132 fast analog to digital converter and stored in a laptop computer. Digitized voltage vs. time data was converted to pressure vs. time using the calibration certificate provided by the manufacturer of the pressure sensor.

Three trials for each design were recorded, and five trials were recorded for the Shchelkin spiral, for it showed the greatest promise. In addition, three trials with the pressure sensor at 20, 40, and 60 mm from the shock tube opening were recorded for the Shchelkin spiral to measure the magnitude of the simulated blast wave as it traveled from the shock tube opening.

3.0 RESULTS

Of the designs tested, the priming section with a Shchelkin spiral inside behind the driving section proved to have the biggest peak pressure. This design achieved the design goal, with an average peak pressure of 5333 kPa (± 98 kPa) while approximating a Friedlander waveform. The bottleneck shock tube with the 103 mm driving section had the second biggest peak pressure, with an average peak pressure of 2642 kPa (± 41 kPa). The bottleneck shock tube with the 79 mm driving section had the third largest peak pressure, with an average peak pressure of 2246 kPa (± 94 kPa). The addition of solid fuel to the driving section had the smallest peak pressure and produced an average peak pressure of 1696 kPa (± 92 kPa). Table 1 shows the average peak pressure, standard deviations, and uncertainties for each of the designs. The uncertainties were calculated as the standard error of the mean. It is of general interest in blast research to know the pressure-time curves generated by each design for comparison with free-field blast waves. The duration of the pressure wave produced by each design varied, from about 0.2 milliseconds for the Shchelkin spiral (Figure 2), to about 2 milliseconds for the other designs (Figures 3, 4, and 5).

*Table 1: The average peak pressures at the shock tube opening for each design, along with standard deviations and uncertainties.*

| Design | Average Peak Pressure (kPa) | Standard Deviation (kPa) | Standard Deviation (%) | Uncertainty (kPa) | Uncertainty (%) |
|---|---|---|---|---|---|
| **Shchelkin Spiral** | 5333 | 219 | 4 | 98 | 2 |
| **79 mm Bottleneck** | 2246 | 162 | 7 | 94 | 4 |
| **103 mm Bottleneck** | 2642 | 70 | 3 | 41 | 2 |
| **Solid Fuel** | 1696 | 156 | 9 | 92 | 5 |

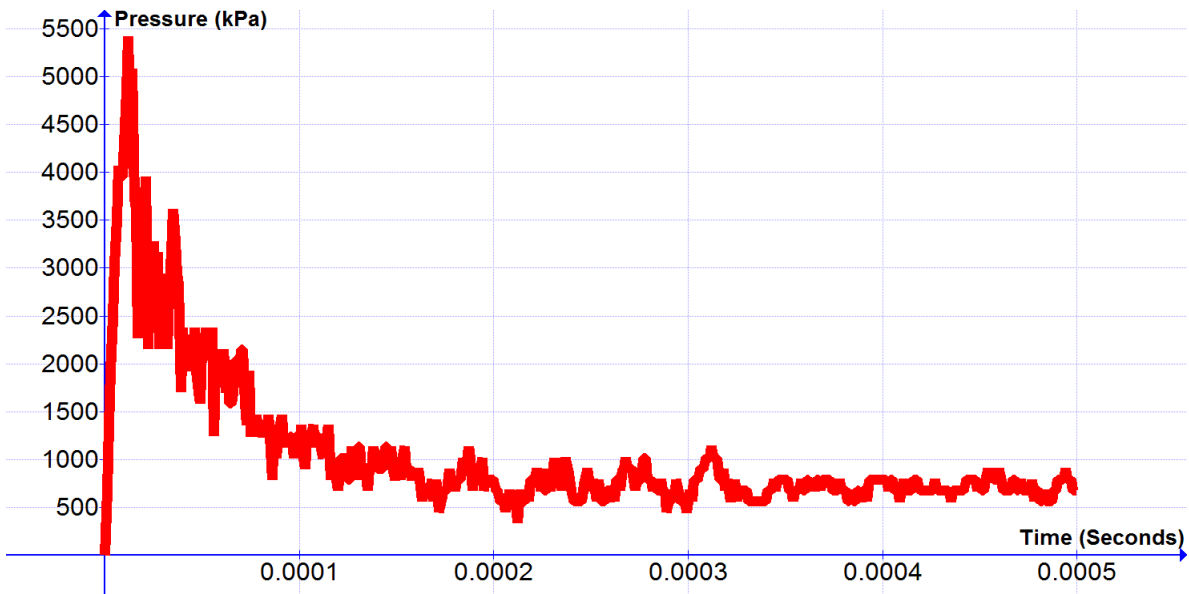

*Figure 2: Blast pressure as a function of time produced by the shock tube with a Shchelkin spiral in the priming section.*

Figure 2 shows pressures measured at the shock tube opening by the pressure sensor with its face perpendicular to the direction of travel of the blast wave at a distance of 0 mm. The shape and magnitude of the blast wave were repeatable. The addition of the Shchelkin spiral resulted in shorter duration blast wave than the base design, because the total energy was similar but condensed into a much shorter time span. Table 2 shows that the peak pressure of the wave created by the design with the Shchelkin spiral does not decay rapidly after leaving the tube. However, the shape of the waveform begins to degrade with distance, suggesting that measurements beyond 60 mm may not be useful due to the degraded wave shape.

*Table 2: Peak blast pressure at distances of 0 mm, 20 mm, 40 mm, and 60 mm from the opening of the Shchelkin spiral shock tube.*

| Distance (mm) | Peak Pressure (kPa) |
|---|---|
| 0 | 5333 |
| 20 | 5493 |
| 40 | 5437 |
| 60 | 5437 |

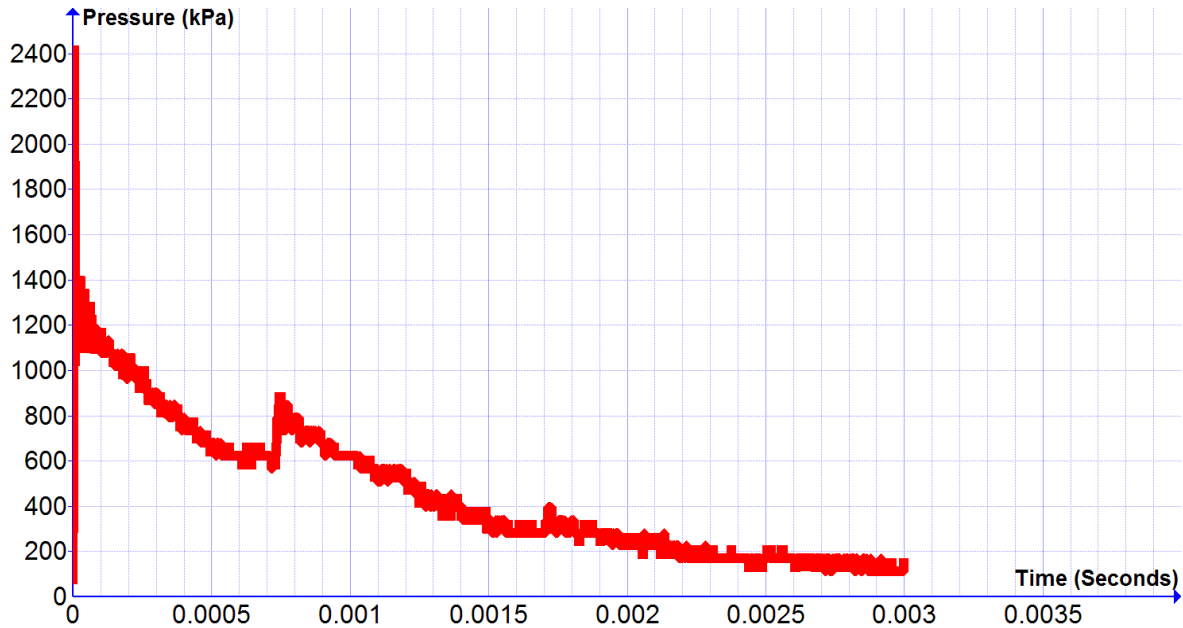

*Figure 3: Blast pressure as a function of time produced by the bottleneck shock tube with a 79 mm inner diameter driving section.*

    Figure 3 shows pressures measured 0 mm from the bottleneck shock tube opening. Note the additional local peaks near 0.75 ms and 1.7 ms. These local peaks are reflections of the blast wave caused by the bottleneck. The shape and magnitude of the blast wave were approximately the same in each of the trials. The bottleneck was used to increase the area of the driving section relative to the driven section to increase the peak blast pressure.

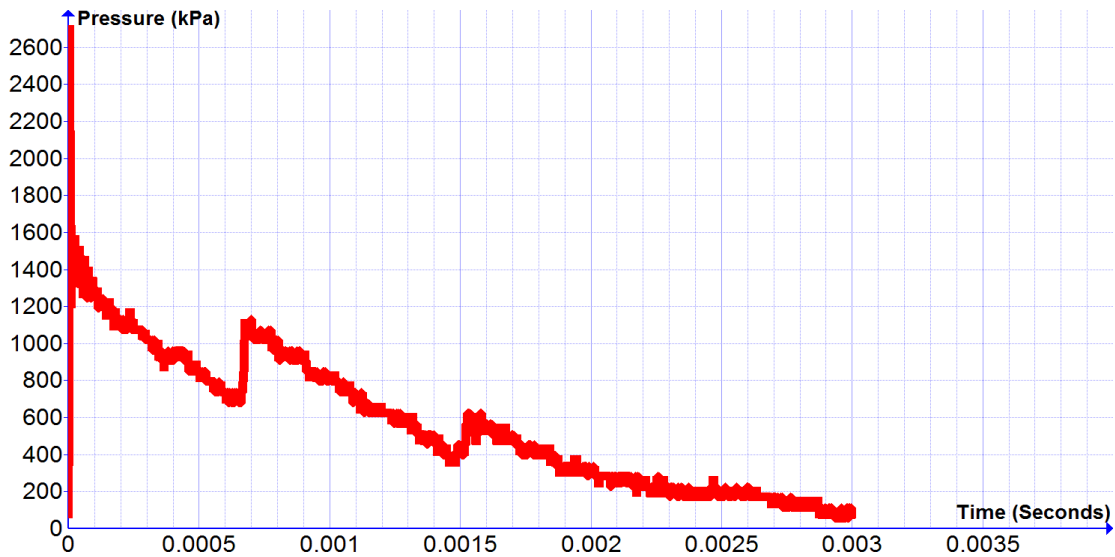

*Figure 4: Blast pressure as a function of time produced by the bottleneck shock tube with a 103 mm inner diameter driving section.*

Figure 4 shows pressures measured 0 mm from the bottleneck shock tube opening. Note the reflections of the blast wave caused by the bottleneck near 0.7 ms and 1.55 ms. The reflections in Figure 4 are more pronounced than the reflections in Figure 3, because the difference in area between the driving section and driven section is larger. The shape and magnitude of the blast wave were about the same in each of the trials. The bottleneck was used to increase the area of the driving section relative to the driven section to increase the peak blast pressure.

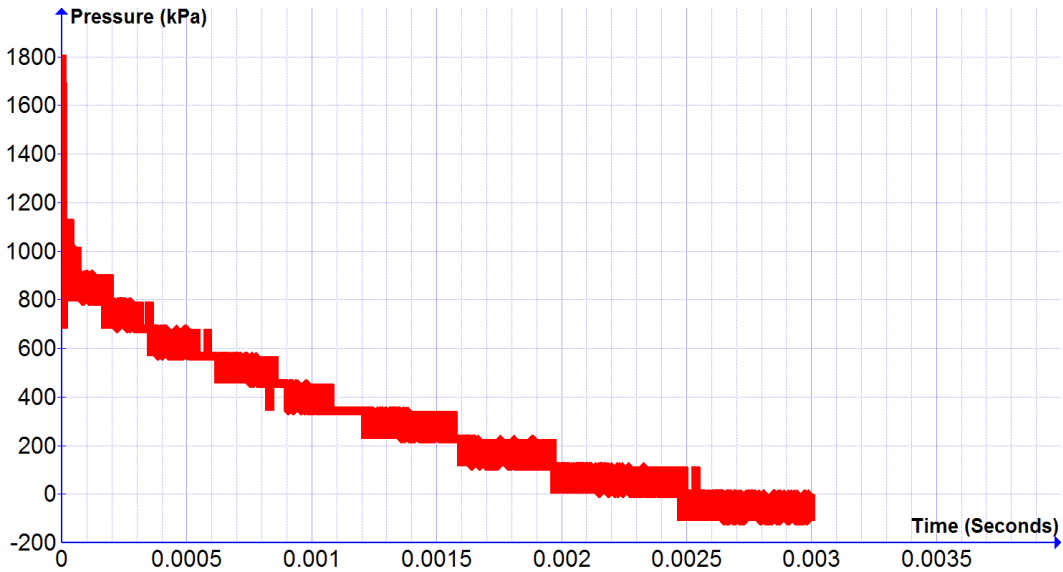

*Figure 5: Blast pressure produced by the solid fuel design as a function of time.*

Figure 5 shows pressure measured at the opening of the solid fuel added design (a distance designated 0 mm). The shape and magnitude of the blast wave were repeatable from trial to trial. The solid fuel was added to increase the available amount of chemical energy for the blast wave.

4.0 DISCUSSION AND CONCLUSIONS

The experimental results showed the differences in the ability of the different designs to increase the peak pressure of a blast wave. The designs had a range of peak pressures from 1.7 MPa to 5.3 MPa. The results from each set of experiments were repeatable, with standard deviations ranging from 3-9%. Another strength is that the shock tube designs require few or no specialized or expensive parts. A single trial in a blast experiment using high explosives may cost above $10,000. In contrast, the designs described here produce realistic simulated blast waves and with materials costs of less than $1000 without the added liability, storage, safety, or personnel costs required when using high explosives.

The blast wave for the Shchelkin spiral design, as shown in Figure 2, is a near-Friedlander waveform, which means that it has a sudden shock front followed by a near-exponential decay. It is not an exact Friedlander waveform because of the "noise" as the shock front decays and due to the absence of a negative phase. This design may be improved by the addition of a short driven section which may smooth out the noisy waveform.

The blast waves for the bottleneck shock tubes are not Friedlander waveforms because the blast wave reflects off the flange used to decrease the surface area (Figure 1B), which causes secondary peaks in the waveform (Figures 3 and 4). These reflections might be reduced by replacing the relatively short bottleneck with a longer, more gradually tapered transition section to reduce the diameter without providing a reflective surface. This shock tube might have the potential to produce a non-ideal blast wave with multiple fronts (as suggested by Cernak and Noble-Haeusslein), "Most shock and blast tubes used in current experimental models replicate the ideal blast wave from an open-air explosion, without the capability to generate a non-ideal blast wave with multiple shock and expansion fronts as seen in real life conditions" [3].

For the bottleneck and solid fuel designs, the peak pressure was expected to increase in direct proportion to the amount of fuel added, based on complete combustion of the fuel. However, that was not observed and the peak pressure was less than would be expected based on proportionality to the chemical energy of the fuel. Alternatively, it is known in physics that the energy of a given wave is proportional to the amplitude squared [16, 17]. Therefore, a more reasonable expectation is that the wave amplitude (in this experiment, pressure) will be proportional to the square root of the energy available. In the 79 mm design, the amount of fuel (and thus the chemical energy) was increased by a factor of 2.22 from the shock tube with a 53 mm inner diameter driving section with an expected increase in peak pressure of a factor of 1.49. The actual increase in peak pressure was by a factor of 1.87. In the 103 mm design, the amount of fuel was increased by a factor of 3.78 and the expected increase in peak pressure was by a factor of 1.94. The actual increase in peak pressure was by a factor of 2.20. For both variations of the bottleneck design, the increase in peak pressure is slightly greater than would be expected if the increase in peak pressure was proportional to the square root of the increase in fuel, but is much smaller than the increase in peak pressure expected if the increase in peak pressure was proportional to the increase in fuel [15].

The waveform generated by the shock tube with solid fuel added (Figure 5) is an approximate Friedlander waveform. For this design, the available energy was tripled from the 53 mm diameter shock tube without solid fuel added [12]. The expected increase in peak pressure would be by a factor of 1.73. The actual peak pressure of the blast wave was only increased by a factor of 1.42. This suggests that not all of the solid fuel was burned quickly enough to contribute to increasing the peak pressure.

Future work may include design improvements such as adding a short driven section to the Shchelkin spiral design to reduce noise in the wave shape or replacing the short bottleneck flange with a longer, gradually tapering section to reduce area and increase peak pressure without reflections. It may be possible to further increase the peak simulated blast pressure by increasing the amount of chemical energy in the Shchelkin spiral or bottleneck designs with a longer driving section. The Shchelkin spiral priming section might also be combined with a longer driving section and/or a gradually tapered transition section to increase the peak pressure above 5 MPa. However, combining Shchelkin spiral or bottleneck designs with the addition of solid fuel is not promising as the solid fuel would have to finish combustion within about 0.2 ms. In the current solid fuel design, the solid fuel has not finished combustion even after 2 ms have passed.

Shock tubes currently employed to test vehicle and structural armor [18] utilize high explosives, compressed gases or specialized equipment. Such tests are expensive and may require getting special permission. However, the results of this experiment demonstrate shock tubes which produced blast waves with higher peak pressures to test vehicle and other armors without requiring specialized equipment or high explosives.

A possible limitation of the shock tube is that the fuel-oxygen ratio initially in the driving sections is not precisely measured. However, consistent procedures were followed, and since the peak pressures and pressure-time curves produced were so consistent from shot to shot, that does not seem to be an important limitation. Although the addition of a Shchelkin spiral did increase the peak pressure to meet the design goal, the reasons why it did so are still unclear. It is also a possibility that the priming compound used for ignition may initiate detonation of the fuel-oxygen mixture without the presence of a Shchelkin spiral. A limitation of the bottleneck design is that the bottleneck produces unrealistic reflections which may have lowered the peak pressure of the initial blast.

*4.1 Conclusions*

In summary, three methods were investigated to increase peak simulated blast pressure produced by an oxy-acetylene driven shock tube while maintaining suitability for laboratory studies. The first method, the addition of a Shchelkin spiral priming section, increased the average peak pressure from 1.17 MPa to 5.33 MPa. The second method was a bottleneck between the driving and driven sections. Coupling a 79 mm diameter driving section to a 53 mm driven section increased the peak pressure from 1.17 MPa to 2.25 MPa. Using a 103 mm driving section increased peak pressure to 2.64 MPa. The third method, adding solid fuel to the driving section with the oxy-acetylene, resulted in a peak pressure increase to 1.70 MPa, and the solid fuel did not completely combust in the available time. Of the new designs tested, the design that incorporated a Shchelkin spiral met the design goal the best and maintained a relevant pressure-time curve (Friedlander waveform).

ACKNOWLEDGMENTS
The authors are grateful to Louisiana Shooters Unlimited for use of their test facilities. This work was funded by BTG Research; affiliated authors participated in all stages of the study with scientific and academic freedom
.